# The Legibility Gap: How Gender Equity Interventions Redistribute Recognition Across Cultures

Binglu Wang[1,2,3,4], Jose Cervantez[5], Jiahui Xue[1,2], Katherine Milkman[5], and Dashun Wang[1,2,3,4,6*]

[1] Kellogg School of Management, Northwestern University; Evanston, IL, USA
[2] Center for Science of Science and Innovation, Northwestern University; Evanston, IL, USA
[3] Northwestern Innovation Institute, Northwestern University; Evanston, IL, USA
[4] Ryan Institute on Complexity, Northwestern University; Evanston, IL, USA
[5] The Wharton School of the University of Pennsylvania; Philadelphia, PA, USA
[6] McCormick School of Engineering, Northwestern University, Evanston, IL, USA

* Corresponding author. Email: dashun.wang@kellogg.northwestern.edu

## Abstract:

Efforts to promote gender equity in science increasingly rely on name-based inference to quantify representation and guide policy and behavior. Yet linguistic cues that signal gender vary across cultures and are often obscured when names are transliterated into English. Here we identify a pattern we call the “legibility gap”: when gender is inferred from names, equity interventions systematically benefit women whose names signal gender while bypassing those whose names lose such cues in translation. Using both observational and experimental evidence, we show how this gap reshapes recognition in science. Analyzing citation diversity statements—an emerging practice in which authors report the algorithmically estimated gender composition of their reference lists—we find that papers that include this practice cite women more frequently, but the gains accrue almost entirely to authors with gender-signaling Western names. By contrast, women whose names lose gender cues in English transliteration, predominantly those with East Asian names, receive fewer citations in these same papers. Two preregistered experiments ($N = 2{,}250$) corroborate this pattern and identify its mechanism: linguistic legibility, not cultural unfamiliarity, determines who is recognized as a woman and who benefits from policies designed to support women in science. Overall, these findings expose a previously unrecognized layer of inequity embedded in global equity infrastructures. As science becomes increasingly global and equity efforts increasingly algorithmic, the legibility gap reveals how uneven identity recognition reshapes fairness. In global systems of recognition, equity depends not only on whether policies are effective on average, but also on whether they are equitable across cultures.

## 1. Introduction

Efforts to promote women in science have accelerated in recent years, in part due to growing awareness of persistent disparities in hiring, credit, promotion, and visibility [1-17]. But promoting equity first requires a way to identify women in the data. Names have long shaped how women are recognized and evaluated [18, 19]: women authors have adopted male pseudonyms to evade bias [20], and controlled experiments show that identical résumés are judged differently depending on whether a male or female name is attached [3, 21-23]. Today, these dynamics operate at a far greater scale. Because verified demographic data are often unavailable, incomplete, or protected for privacy reasons [24], researchers and institutions increasingly rely on name-based inference—by algorithms and by human readers—to identify gender and monitor inclusion efforts [5, 6, 8, 9, 12-16].

Yet names are not equally legible across languages. In many naming systems, gender cues are embedded in spelling, characters, or pronunciation that disappear when names are transliterated into English [25-27], the dominant language of science [28, 29]. As a result, some women's names remain readily recognizable as female while others do not, even when gender is clear in the original language. When equity efforts depend on inferred gender, this uneven legibility may create a consequential asymmetry: the very systems designed to support women may end up supporting only those whose names are easiest to classify as women.

While name-based inference rarely determines an outcome on its own, its significance lies upstream at the opportunity layer where equity efforts typically intervene, by shaping who is visible in a candidate or expert pool, how representation is measured, which disparities are diagnosed, and where corrective attention is directed (Table S1). This infrastructure now operates across hiring and workforce audits, official statistics, expert directories, and large-scale analyses of science [5, 6, 8, 9, 12-16] (SI Section S1). Recent work, including our validation across 4.5 million scientists (see SI Section S1.1), shows that these systems perform substantially worse on transliterated East Asian names than on Western ones [25, 26, 30]. Uneven legibility may therefore be more than a source of measurement error. By shaping who is counted, surfaced, and targeted by equity efforts, it may redistribute opportunity toward those whose identities are easiest for the system to see.

In this paper, we test that possibility using both observational and experimental evidence. We first examine citation diversity statements, an emerging practice in which authors report the gender composition of their reference lists, creating a natural setting in which inferred gender shapes behavior [31, 32]. We then present two preregistered experiments involving participants from five cultures that test whether interventions designed to support women in science disproportionately benefit those whose names remain legible as female in English. Across both settings, we find that interventions intended to promote gender equity help women overall but disproportionately benefit women with Western names that clearly signal gender, while bypassing women whose names lose gender cues in transliteration. We term this pattern the "legibility gap."

## 2. Study 1: Citation diversity statements and the redistribution of recognition

To test whether equity efforts built on inferred gender may redistribute recognition unevenly across cultures, we begin by analyzing citation diversity statements (CDSs) in science. CDSs make inferred gender behaviorally consequential: authors are encouraged to calculate the gender composition of the scholars they cite using an automated classification tool and then disclose that composition in their published paper [31, 32]. If gender cues are not equally legible across names, then a practice designed to increase citations to women may not benefit all women equally. Instead, it may disproportionately reward those whose names are easiest for the system to recognize as female.

As illustrated in Fig. 1a, we assembled a dataset of papers published between 2020 and 2024[1] that included a CDS and compared them with matched papers from the same journals that did not. After matching on publication characteristics and excluding papers without reference records, the final sample includes 216 papers with CDSs and 432 matched papers (total N=648 papers), together citing 40,719 references. For each cited author across these 648 papers, we inferred gender and cultural origin from the name, allowing us to examine whether papers adopting CDSs differ from their matched comparison papers not only in how often they cite women, but also in which women benefit from those shifts in citation practice. We estimate a series of paper-level regressions across matched papers where the outcomes of interest are various measures of citation composition and the primary predictor is always an indicator for the inclusion of a CDS, and where we control for publication year, field, region, team size, and citation impact (see SI Section S2 for details).

[1] The dataset was collected in May 2024.

We find that papers with citation diversity statements cite more women than matched papers without them. The share of women's names cited is 7.0 percentage points higher in papers with CDSs than in matched comparisons (30.8% vs. 23.8%; Fig. 1b), and this difference remains substantial after controlling for publication year, field, region, team size, and citation impact ($\beta = 0.0709, p < 0.001$; Table S2). However, that gain is not evenly distributed. The same papers show almost no differences in citations to men's names ($\beta = 0.0108, p = 0.297$), but a much sharper gap in citations to gender-blind names ($\beta = -0.0860, p < 0.001$). This suggests the apparent improvement does not reflect a simple redistribution from men to women, but a shift away from names whose gender is less legible to the system.

This asymmetry becomes sharper when we examine the cultural origin of cited names. Papers with CDSs cite significantly more authors with English-origin names ($\beta = 0.0606, p < 0.001$), but fewer authors with Chinese names ($\beta = -0.0316, p < 0.001$), Japanese names ($\beta = -0.0101, p < 0.001$), and Arab names ($\beta = -0.0080, p < 0.001$), among others (Fig. 1c, Table S3). In other words, although CDSs are associated with increased citation of women overall, those gains accrue disproportionately to women whose names remain legible as female to English-oriented inference systems. And people whose names lose gender cues in transliteration do not share equally in the benefits of the intervention.

Overall, these results reveal a paradox at the heart of name-based equity interventions. A practice intended to make citation behavior more gender equitable is associated with improvements in recognition for women overall, but it is also associated with more recognition for women whose names are easiest to classify as female. Because this evidence comes from observational data, however, it cannot by itself establish that the intervention causes this asymmetry, nor can it distinguish linguistic legibility from a number of alternative explanations for the detected relationships. We therefore turn next to preregistered experiments that isolate the causal effect of gender-equity interventions and test whether the same asymmetry emerges under controlled conditions.

## 3. Study 2: A causal test of the legibility gap

To isolate the causal effect of equity-promoting interventions, we conducted a preregistered experiment (https://aspredicted.org/py94-n9g5.pdf) in which participants selected scholars for a real Facebook campaign promoting outstanding academics. We tested whether motivating people to increase gender diversity disproportionately benefits women with Western names over those with Eastern names. As illustrated in Fig. 2a, we recruited 750 participants from the United States and asked each to choose scholars from a list containing both men and women from Western and Eastern countries to include in the advertisement. After making an initial set of selections, participants were randomly assigned to one of two conditions. In the control condition, they received feedback summarizing non-gender attributes of their selections. In the treatment condition, they were instead told what percentage of their selected scholars were women, following an established design for boosting the selection of women [33]. All participants then chose one additional scholar, allowing us to test whether gender feedback changed who benefited from the intervention (see SI Section S3 for full stimuli and design details). The key question was whether feedback designed to increase gender diversity would benefit all women equally, or primarily those whose names remain clearly legible as female in English.

Following our pre-registered plan, we find that gender feedback increased the selection of women, consistent with prior studies [33]. Yet importantly, here we find that this gender feedback did not benefit all women equally. Participants randomly assigned to receive gender feedback were more likely to choose a Western woman as their final pick than participants in the control condition (21.1% versus 10.8%; $\beta = 0.1029, p < 0.001$; Fig. 2b, Table 1 Model 1). By contrast, the same feedback had no statistically significant effect on the likelihood of selecting an Eastern woman (10.8% versus 8.2%; $\beta = 0.0265, p = 0.217$; Fig. 2b, Table 1 Model 3). This difference remains substantial after controlling for participants' demographics (Table 1 Model 2 and 4). A pre-registered Wald test comparing these treatment effects controlling for demographics confirms that the intervention benefited Western women significantly more than Eastern women ($\chi^2(1) = 4.450$ , $p = 0.0349$).

Study 2 thus provides causal evidence for the pattern documented by Study 1. An intervention designed to increase women's representation does increase women's selection overall, but the benefit is concentrated among women whose names more clearly signal female identity in English. One possible explanation, however, is that this asymmetry reflects the cultural background of the evaluators themselves: perhaps U.S. participants respond more readily to Western names because those names are more familiar to them. We therefore turn to a second preregistered experiment designed to distinguish linguistic legibility from cultural familiarity.

# 4. Study 3: Testing whether shared cultural background closes the legibility gap

If the legibility gap primarily reflects cultural unfamiliarity, then it should weaken when evaluators share the relevant naming context. To test this possibility, our second preregistered experiment recruited participants from both Western and Eastern countries and tested whether evaluators were more responsive to the intervention when judging women from their own broader cultural background. This design allows us to distinguish two mechanisms: a familiarity account, in which shared cultural context should reduce the asymmetry, and a legibility account, in which the asymmetry should persist even across evaluator groups. This study was pre-registered on OSF (https://osf.io/j2ksg).

Here, we recruited 1,500 participants from China, South Korea, Italy, and Germany (375 from each) for a task that closely paralleled Study 2. Each participant selected scholars for a Facebook campaign from a pool containing candidates from all four countries and then received either feedback about the gender distribution of their initial selectees or feedback about other features of their initial selectees before making their final selection. As in Study 2, the primary outcome was whether the final selected scholar was a woman from a Western or Eastern country (see SI Section S4.1 for details).

We first replicated the basic asymmetry from Study 2. Among 1,443 participants who completed the task, gender feedback substantially increased the selection of Western women, from 7.8% in the control condition to 21.4% in the treatment condition ($\beta = 0.1360, p < 0.001$; Table 2 Model 1). By contrast, it produced no statistically significant increase in the selection of Eastern women ($\beta = 0.0184, p = 0.233$; Table 2 Model 3). A pre-registered Wald test comparing these two coefficients confirms that the difference in treatment effects is statistically significant ($\chi^2(1) = 21.157, p < 0.001$). Thus, even in a more culturally heterogeneous sample, the intervention again benefited Western women far more than Eastern women.

Crucially, this pattern held even when evaluators shared their broad cultural background with the women they were evaluating. Among Western participants, gender feedback increased the selection of Western women from 7.8% to 25.3%; among Eastern participants, it increased the selection of Western women from 7.8% to 17.7%. However, the same feedback had little effect on selections of Eastern women by either participant group: among Western participants, selection rose only from 9.2% to 11.5%, and among Eastern participants, it only rose from 8.1% to 9.0% (Fig. 3). An interaction analysis shows that Western participants responded somewhat more strongly than Eastern participants to gender feedback when selecting Western women ($\beta_{Western_woman:\ feedback \times western_participant} = 0.0750, p = 0.041$; Table 2 Model 2), but no comparable difference between participant groups emerged when selecting Eastern women ($\beta_{Eastern_woman: feedback \times western_participant} = 0.0188, p = 0.543$; Table 2 Model 4). A pre-registered Wald test comparing these two interaction terms finds they do not differ significantly ($\chi^2(1) = 1.204, p = 0.272$), providing no evidence that participant region differentially moderates the feedback effect on Western- versus Eastern-woman selection. In follow-up analyses where we examined the decisions by Chinese, South Korean, Italian and German participants about women from their own culture, separately, we find exactly the same pattern of results (SI Section S4.2 and Table S5).

Study 3 therefore sharpens the mechanism underlying the legibility gap. The asymmetry is not simply a byproduct of Western evaluators being unfamiliar with Eastern names. Even evaluators from the same Eastern countries conferred substantially more benefit to Western women than to Eastern women when exposed to a nudge designed to promote gender diversity. Hence, what matters is not merely who is judging, but whether female identity remains legible in the names on which the intervention depends. Together, Studies 2 and 3 show that the unequal gains documented in Study 1 are consistent with benefits that accrue causally: when policies rely on inferred gender, linguistic legibility determines who is recognized as a woman and who benefits from efforts to support women in science.

## 5. Discussion

Our findings reveal a structural paradox in contemporary efforts to promote gender equity in science. Interventions designed to increase women's recognition do not benefit all women equally when they rely on names to infer gender. Instead, they disproportionately advantage women whose names remain legible as female in English, while bypassing women whose gender cues are obscured in transliteration. Across observational and experimental evidence, we show that progress toward gender parity can therefore come with an unrecognized tradeoff: gains in average inclusion may mask a redistribution of recognition across linguistic and cultural lines.

Name-based inference approaches make efforts to measure and reduce potential gender gaps more scalable and feasible across large datasets and organizations, and they have become a key part of the infrastructure through which institutions measure representation, diagnose inequity, and design corrective interventions. In that context, errors in gender inference are not merely measurement problems [34-36]. They can become allocation problems. When some identities are more legible than others, the very tools designed to improve fairness can channel recognition toward those whose identities are most legible to the system. Our results thus shift attention from asking whether equity interventions work on average to addressing a deeper question: for whom do they work, and under what conditions?

Our experiments clarify the mechanism underlying this asymmetry. The unequal gains we document are not simply the product of Western evaluators favoring culturally familiar names. Even when evaluators come from Eastern countries, interventions intended to increase women's representation continued to benefit Western women far more than Eastern women. What matters, then, is not just who is judging, but whether female identity remains legible in the names on which judgment depends. In this sense, the legibility gap is interpreter-general: it emerges in both algorithmic and human inference when names function as the proxy through which gender is encoded.

Our findings also illuminate an overlooked tension in the current turn toward data-driven equity [37]. Over the past decade, scholars, journals, and institutions have increasingly embraced inclusive citation practices, equity dashboards, and algorithmic auditing tools to monitor representation and correct disparities. Here, our results show that when inclusion is operationalized through proxies such as names, the universality of measurement can conceal the cultural specificity of the signals being measured. A system may appear neutral because it applies the same rule to everyone, while still systematically favoring those whose identities are more easily legible within the linguistic conventions on which that rule depends [38, 39].

The broader implication extends beyond gender classification. In many global evaluative settings, recognition depends on signals that must travel across languages, scripts, and cultural contexts [40-42]. When those signals survive unevenly, systems may reward what remains easiest to interpret [43]. Legibility, in this sense, represents a hidden dimension in evaluation systems. The question is not just who is represented, but whose identity remains legible enough to be counted in the first place. This suggests a broader agenda for studying fairness in transnational and algorithmically mediated settings.

Several limitations qualify our analysis and point to important future directions. Our focus on the distinction between specific "Western" and "Eastern" countries' naming systems is somewhat coarse, and gender legibility likely varies within countries, across diasporas, and across hybrid naming practices. Our analyses also inherit the binary gender assumptions built into existing inference tools, leaving open important questions about nonbinary identities and culturally specific gender systems [44]. Finally, we focus on immediate outcomes in citation and selection; future work should examine whether legibility-based advantages compound over time through cumulative processes of visibility, collaboration, and advancement.

Taken together, our results highlight a fundamental challenge for equity in an increasingly global and algorithmic scientific system. Inclusion depends not only on whether interventions expand opportunity on average, but also on whether the categories through which opportunity is measured remain legible across languages and cultures. The legibility gap shows how infrastructures designed to enhance fairness can silently reproduce the hierarchies of the languages that feed them. In global systems of recognition, equity depends not only on equal opportunity, but on equal legibility.

**Data and Code Availability**: All analysis code, data files, and an extended appendix containing additional tables will be made available.

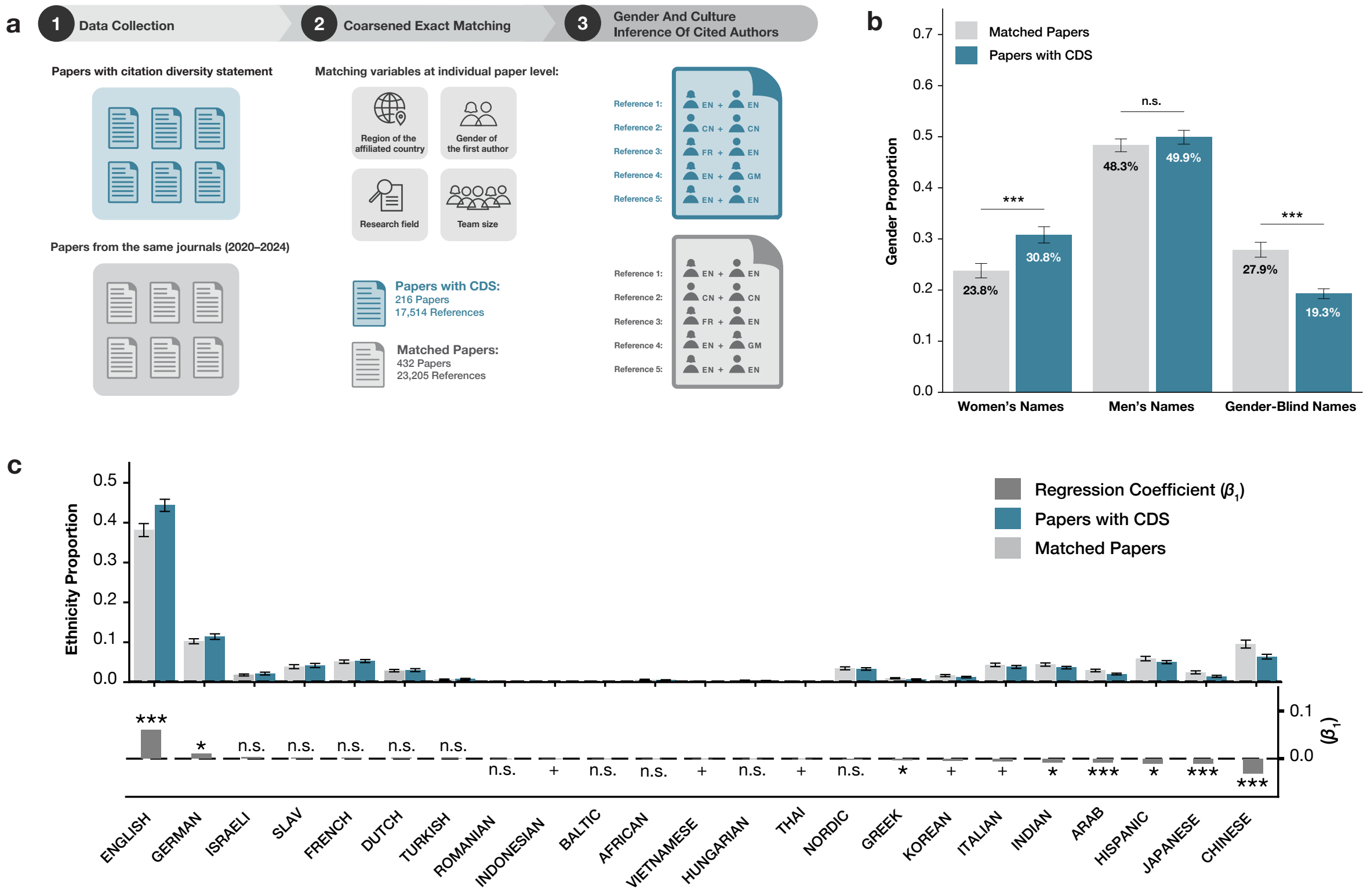

**Figure 1. Citation diversity statements and the redistribution of recognition.** (a) We identified 239 research articles containing a citation diversity statement and assembled a matched comparison group using Coarsened Exact Matching (CEM), yielding 216 papers with a CDS and 432 matched papers. For each paper, we extracted the first- and last-author names of their references and inferred (1) gender using the *gender-guesser* library and (2) cultural origin using *Ethnea*. (b) Gender composition of cited authors. Papers with citation diversity statements cited women (30.8%) at a rate 7.0 percentage points higher than the matched sample (23.8%). This difference was offset by a significantly lower (8.6 percentage points lower) rate of citations to gender-blind names, and there was no corresponding difference in citations to men, suggesting a preference for names that clearly signal a female identity. (c) Cultural composition of cited authors' names. Papers with citation diversity statements cite significantly more authors with English-origin names, while citing several non-Western cultural groups less frequently. Bars display means ± 1 standard error and are ordered by regression coefficients. $+p < 0.10, * p < 0.05, ** p < 0.01, *** p < 0.001$.

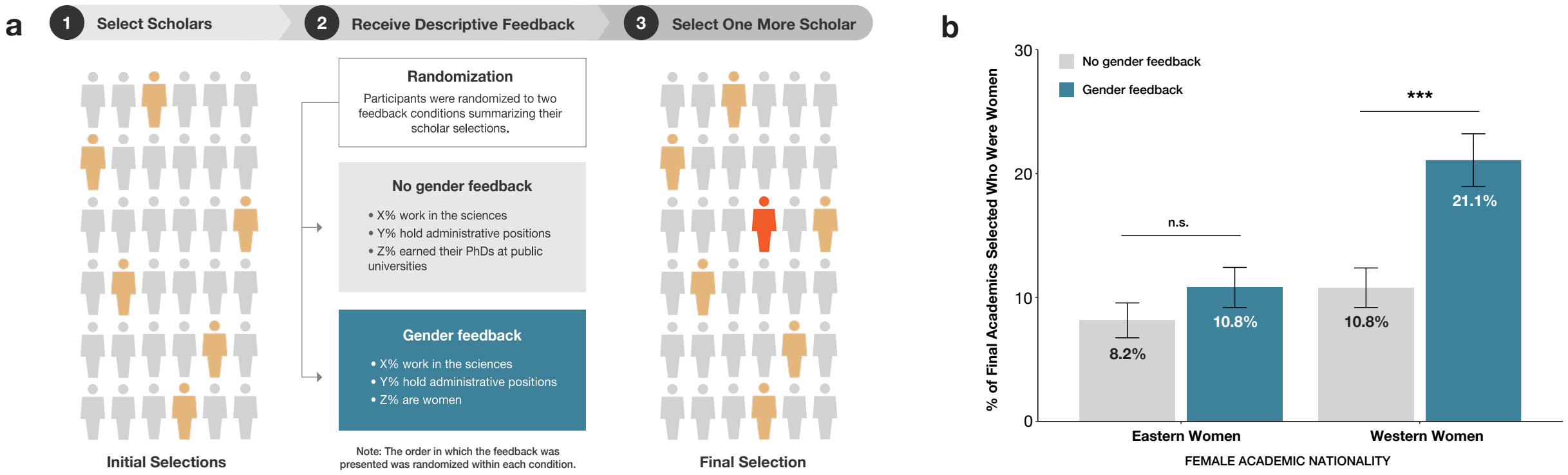

**Figure 2. The effect of gender feedback on the selection of a seventh female scholar, split by scholar nationality (Study 2).** (a) This panel illustrates the experimental design workflow. Participants were first presented with a list of scholars from which they selected six to feature in a Facebook advertisement about academics. In Step 2, participants were randomly assigned to one of two feedback conditions, each providing a summary of the characteristics of the scholars they had initially selected. In the gender-feedback condition, participants received feedback summarizing the gender composition of their six initial selections; in the no-gender-feedback condition, no gender feedback was provided. In Step 3, participants were asked to select a seventh scholar, and the gender identity of their final selectee is our primary dependent measure. (b) The share of participants who chose a female scholar as their final (seventh) pick, disaggregated by whether the scholar was from an Eastern or Western background and by experimental condition (gender-feedback vs. no-gender-feedback). The x-axis distinguishes scholars from Eastern countries (China, South Korea, Thailand) and Western countries (United Kingdom, Italy, Germany). Grey bars represent the no-gender-feedback condition, in which participants received no feedback summarizing the gender mix of their initial six selections; blue bars represent the gender-feedback condition, in which participants were given feedback summarizing the gender composition of their initial six selections before picking a seventh. Error bars denote ± 1 standard error. $+p < 0.1$; $*\ p < 0.05$; $**\ p < 0.01$; $***\ p < 0.001$.

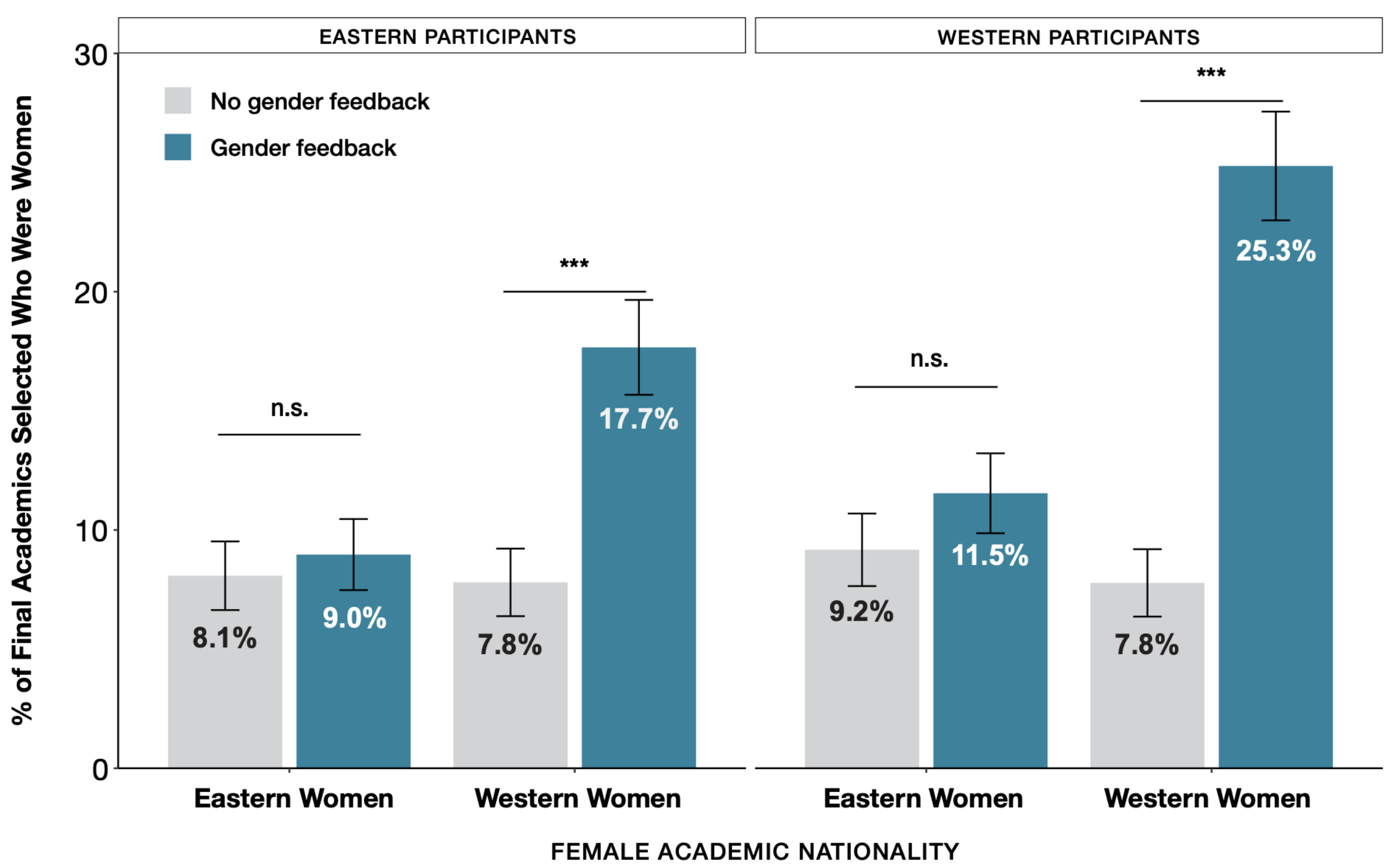

**Figure 3. The effect of gender feedback on the selection of a seventh female scholar, split by participant region and scholar nationality (Study 3).** The chart shows the share of participants who chose a female scholar as their final (seventh) pick in Study 3, disaggregated by whether the scholar was from an Eastern or Western background and by experimental condition (gender feedback intervention vs. control). The Left panel shows data from participants from Eastern countries (China, South Korea), and the Right panel shows data from participants from Western countries (Germany, Italy). Within each panel, the x-axis distinguishes between selected scholars from Eastern nations (China or South Korea) and Western nations (Germany or Italy). Grey bars represent the *no-gender-feedback* condition, in which participants received no feedback summarizing the gender mix of their initial six selections; blue bars represent the *gender-feedback* condition, in which participants were given feedback summarizing the gender composition of their initial six selections before picking a seventh scholar. Error bars denote ± 1 standard error. $+p < 0.1$; $*\ p < 0.05$; $**\ p < 0.01$; $***\ p < 0.001$.

| Dependent variable: Identity of final academic selected: | Western Woman | | Eastern Woman | |
|---|---|---|---|---|
| | Model 1 | Model 2 | Model 3 | Model 4 |
| **Intervention** (Feedback was provided on the % of academics who were women) | 0.1029***<br>(0.0266) | 0.1037***<br>(0.0267) | 0.0265<br>(0.0215) | 0.0268<br>(0.0216) |
| Controls | No | Yes | No | Yes |
| Observations | 750 | 750 | 750 | 750 |
| R-Squared | 0.0198 | 0.0259 | 0.0021 | 0.0030 |
| Wald Test (Intervention in Model 2 vs. Model 4) | $\chi^2(1) = 4.450$ , $p = 0.0349$ | | | |

**Table 1. OLS Regressions estimating the effect of gender feedback on participants' final selection (Study 2).** The table presents four OLS regressions predicting whether the seventh (final) scholar chosen by a participant for inclusion in a Facebook advertisement was a Western woman (Models 1–2) or an Eastern woman (Models 3–4). The key independent variable in every model is a binary indicator for random assignment to receive feedback on the percentage of women in the participant's initial six-scholar roster. Models 2 and 4 add demographic controls: participant gender (male = 1), race (White = 1), and age. The Wald test row reports a cross-equation test comparing the Intervention coefficient across Model 2 and Model 4. Robust (HC3) standard errors are shown in parentheses. $+p < 0.1$; $* \, p < 0.05$; $** \, p < 0.01$; $*** \, p < 0.001$.

| Dependent variable: Identity of final academic selected: | Western Woman | | Eastern Woman | |
|---|---|---|---|---|
| | Model 1 | Model 2 | Model 3 | Model 4 |
| **Intervention (**Feedback was provided on the % of academics who were women) | 0.1360*** (0.0183) | 0.0985*** (0.0259) | 0.0184 (0.0154) | 0.0090 (0.0218) |
| **Intervention x** Western Participant | | 0.0750* (0.0366) | | 0.0188 (0.0308) |
| Western Participant | 0.0066 (0.0341) | -0.0323 (0.0390) | 0.0303 (0.0287) | 0.0205 (0.0329) |
| Controls | Yes | Yes | Yes | Yes |
| Observations | 1,443 | 1,443 | 1,443 | 1,443 |
| R-Squared | 0.0413 | 0.0441 | 0.0034 | 0.0036 |
| Wald Test (Intervention in Model 1 vs. Model 3) | $\chi^2(1) = 21.157, p < 0.001$ | | | |
| Wald Test (Intervention x Western Participant in Model 2 vs. Model 4) | $\chi^2(1) = 1.204, p = 0.272$ | | | |

**Table 2**. **OLS Regressions estimating the effect of gender feedback on participants' final selection (Study 3).** The table presents four OLS regressions predicting whether the seventh (final) scholar chosen by a participant for inclusion in a Facebook advertisement was a Western woman (Models 1–2) or an Eastern woman (Models 3–4). The key independent variable in every model is a binary indicator for random assignment to receive feedback on the percentage of women in the participant's initial six-scholar roster. Models 2 and 4 add an interaction with the participant's region of residence (Western = Germany/Italy; Eastern = China/South Korea). All models include demographic controls and incentive fixed effects: participant gender (male = 1), race (White = 1), and age. After excluding participants who did not complete the experiment, the analytic sample comprises 1,443 of the 1,500 recruited participants. The Wald test rows report cross-equation tests comparing the Intervention coefficient across Model 1 and Model 3, and the Intervention x Western Participant coefficient across Model 2 and Model 4. Robust (HC3) standard errors appear in parentheses. $+ p < 0.1$; $* \ p < 0.05$; $** \ p < 0.01$; $*** \ p < 0.001$.

Supplementary Information for

# The Legibility Gap: How Gender Equity Interventions Redistribute Recognition Across Cultures

Binglu Wang, Jose Cervantez, Jiahui Xue, Katherine Milkman, and Dashun Wang*

* Corresponding author. Email: dashun.wang@kellogg.northwestern.edu

# Table of Contents

The sections below are organized to mirror the structure of the main text, moving from the observational foundation in Study 1 through the two preregistered experiments in Studies 2 and 3. Section S1 validates the name-based gender- and cultural-origin inference tools across the 4.5 million scientists in the SciSciNet database and documents the cross-cultural legibility gap (robustness Figs. S1-S5). Section S2 details the Study 1 citation diversity statement (CDS) analysis (main-text Fig. 1; Tables S2-S3). Sections S3 and S4 detail the two preregistered experiments (Study 2 and Study 3) that test the gap causally (main-text Figs. 2-3 and Tables 1-2; Table S5). Supporting figures and statistical tables appear in the Figure and Table Appendices (Sections S5 & S6).

## 1. Gender and Culture-Origin Inference from Names

In many evaluative settings, name-based gender inference functions as a de facto infrastructure for measuring and managing representation. Organizations rarely allocate opportunities on the basis of names alone. Instead, when verified or self-reported demographic data are unavailable, incomplete, or restricted, name-based inference operates upstream: it determines how representation is measured, which disparities become visible, and where corrective attention is directed. For example, talent analytics platforms use inferred gender to characterize candidate and workforce pools; firms and public agencies use it to audit representation and produce statistics on inventors and scientists; and expert-search systems use demographic information to assess or shape who is surfaced for visibility and opportunity (Table S1). Name inference therefore matters not because it is typically the sole basis for a decision, but because it helps determine who is counted, which gaps are seen, and where interventions are aimed.

Science is a particularly consequential setting in which these uses converge. Major bibliometric databases, including Scopus, OpenAlex, and Web of Science, do not contain verified gender fields. Consequently, large-scale studies, dashboards, and policy reports on gender in science commonly infer gender from researchers' names. This dependence is especially consequential because science is both highly global and increasingly governed through scalable data infrastructures: researchers from many naming cultures are compared within the same databases, while citation audits, representation metrics, and expert searches often have only names as a portable identity signal across records and national boundaries.

Yet that signal is not equally informative across naming cultures. Prior accuracy audits, together with our validation across 4.5 million scientists, show that gender-inference systems perform substantially worse on transliterated East Asian names than on Western names. Some large-scale studies respond to this limitation by excluding East Asian names or treating them as unclassified. Although such practices may reduce classification error, they create a substantive blind spot: the scientists least legible to inference systems become least visible in the evidence used to diagnose and address inequality. The analyses below therefore first establish how gender legibility varies across naming cultures and then describe the gender- and cultural-origin inference procedures used in our studies.

### 1.1.Cross-cultural disparities in the gender information conveyed by names

A name reveals a person's gender only when its language of origin encodes that signal in a way that survives transliteration into English. For many Western names, it does; for many East Asian names, it does not. Recent accuracy audits of major gender-inference tools underscore this problem: precision rates drop sharply for transliterated East Asian names compared to Western names [1-3]. Here we extend these audits to a far larger, more policy-relevant and representative population: 4.5 million scientists whose names appear in the SciSciNet database [4], which constitutes nearly the entire global scientific workforce from 1940 to 2022. This large-scale analysis allows us to quantify the magnitude and structure of cross-cultural disparities in the gender information that names convey.

We infer gender and cultural origin from each scholar's name using two widely used open-source tools [5-9]: gender-guesser (https://pypi.org/project/gender-guesser/) for gender inference and Ethnea [10] for cultural origin inference (see Section S1.2 and S1.3 for full tool descriptions). We define gender-blind names as those that gender-guesser classifies as Androgynous (internally coded as "andy") or unknown, and gender-signaling names as those it classifies as female, male, mostly female, or mostly male. This classification provides a quantitative measure of how reliably gender can be inferred from linguistic form alone. As shown in Fig. S1, gender legibility varies dramatically across cultures. Gender is readily inferable from Italian, English, German, and Nordic names, where gender-guesser identifies a gender for around 90% of these names. However, gender is substantially harder to infer from Thai, Indonesian, Korean, and Chinese names, for which the tool recovers gender for only about 10%. The difference in the rate at which gender can be inferred from Western versus Eastern names is statistically significant ($p < 0.001$), confirming that Western names are predominantly gender-signaling, whereas Eastern names are more frequently gender-blind.

To ensure these findings do not depend on any single classifier, we repeated the analysis using several additional tools: two commercial APIs (Genderize.io and Nationalize.io), the Genni companion gender model paired with Ethnea, and a cultural-consensus gender probability from SciSciNet (P(gf)). All four gender tools produce substantively similar cross-cultural patterns (Figs. S1-S4), and the legibility gap persists when cultural origin is assigned by Nationalize.io instead of Ethnea (Fig. S5). Below we describe each tool's origin, input, output, and intended use in this study. Table S4 provides a summary overview.

### 1.2. Gender Inference Tools

The gender-guesser Python library (version 0.4.0; https://pypi.python.org/pypi/gender-guesser/) predicts gender from first names using a manually curated name list. Given a first-name string, it returns one of six categories: female, male, mostly_female, mostly_male, andy (roughly equal use across genders), or unknown (name not found). We passed each author's first name through the library after stripping diacritics and normalizing case. No additional parameters or training data are required; the underlying name list was manually validated by native speakers. gender-guesser is the primary tool used in the main text. As shown in Fig. S1, it classifies the vast majority of Italian, English, German, Nordic, and other European names with high confidence, while returning unknown for the majority of Chinese, Thai, Korean, and Indonesian names.

Genni is the companion gender predictor to Ethnea [11], designed to incorporate Ethnea's ethnicity context to improve predictions for names with gender associations that vary across cultures. Genni takes a first name (optionally with surname or country context) and returns M (male), F (female), or "–" (unknown). We accessed Genni via the same web service as Ethnea (September 2024). The cross-cultural legibility pattern produced by Genni (Fig. S2) closely mirrors that of gender-guesser.

Genderize.io (https://genderize.io/), a commercial API provided by Demografix ApS, draws on a database of over one billion name records aggregated from online sources across many countries. Given a first name, it returns a predicted gender (male or female), a probability between 0 and 1, and a count of supporting records; probabilities near 0.5 indicate unisex names. We queried the API in September 2024 for each unique first name in our author list, applied no country constraint, and assigned each name the gender with the higher probability. As with gender-guesser and Genni, Genderize.io produces high-confidence classifications for Western names and substantially lower recognition rates for Eastern names (Fig. S3).

As a fourth approach, we used gender probabilities from the SciSciNet_Authors_Gender dataset [4], which we refer to as SciSciNet P(gf). This dataset applies the cultural-consensus model of Van Buskirk et al. [12], which combines 36 name-gender sources (national statistics, name lists, etc.) spanning over 150 countries to estimate the probability P(gf) that a given first name corresponds to a female individual, together with an uncertainty measure. We accessed the data in October 2024 and matched each author to their P(gf) via SciSciNet's AuthorId. Authors recorded with initials only (~23% of SciSciNet) cannot be assigned a P(gf) value.

Together, these four tools yield convergent evidence that our gender-inference findings are not artifacts of any single method. That said, all name-based inference tools have limits and can introduce error or bias for any given individual, particularly for names that cross linguistic or cultural boundaries. We therefore interpret our results at the population level and avoid drawing conclusions about any specific individual's gender.

### 1.3. Cultural-Origin Inference Tools

Ethnea is an instance-based classifier developed by Torvik and Agarwal [10] that assigns probable ethnic origin to names. Given a full name, Ethnea retrieves all instances of the name in a large bibliographic database (e.g., PubMed Authority) together with their geocoded affiliations, and probabilistically assigns the name to predefined ethnicity categories (e.g., English, Chinese, Hispanic, Arab). When no single ethnicity dominates or data are insufficient, the output is UNKNOWN; names too short to classify are marked TOOSHORT. We submitted each author's full name to the Ethnea web service (http://abel.lis.illinois.edu/cgi-bin/ethnea/search.py, accessed September 2024) with default settings, and used the returned labels for the visualizations and analyses in main-text Fig. 1c and Table S3.

Nationalize.io (https://nationalize.io/), the commercial counterpart to Genderize.io (also provided by Demografix ApS), predicts the likely country (or countries) of origin for a given surname, returning a ranked list of country codes with probabilities summing to 1. We submitted each author's surname in September 2024 and recorded the top-ranked country and its probability. To verify that the cross-cultural gender-inference pattern is not an artifact of our cultural-origin classifier choice, we replicated the Fig. S1 analysis using Nationalize.io to assign cultural groups; the legibility gap persists (Fig. S5).

Because Genderize.io and Nationalize.io are proprietary services, their underlying datasets are not fully transparent and are updated over time. We report access dates for reproducibility and treat these APIs as cross-checks rather than as ground truth.

## 2. Study 1 Method

Having validated the name-based inference tools and documented the legibility gap at the population level, we now turn to Study 1, which examines whether this gap shapes citation practices in real academic papers. This section details the data and analysis behind Study 1 (main-text Fig. 1; Tables S2-S3): how we identified papers containing a citation diversity statement (CDS), constructed the matched comparison group, and specified the regressions.

### 2.1. Data Collection

To identify papers with a Citation Diversity Statement (CDS), we anchored our search on the cleanBib community template (https://github.com/dalejn/cleanBib), which provides standardized language widely adopted by CDS-using authors. Two sentences from this template appear consistently in published statements:

- "Here we sought to proactively consider choosing references that reflect the diversity of the field in thought, form of contribution, gender, race, ethnicity, and other factors."
- "First, we obtained the predicted gender of the first and last author of each reference by using databases that store the probability of a first name being associated with a woman."

The template cites two anchor references [13, 14] that most CDS-adopting papers also cite:

- J. D. Dworkin, K. A. Linn, E. G. Teich, P. Zurn, R. T. Shinohara, and D. S. Bassett, "The extent and drivers of gender imbalance in neuroscience reference lists," Nature Neuroscience, 2020.
- D. Zhou, M. A. Bertolero, J. Stiso, E. J. Cornblath, E. G. Teich, A. S. Blevins, Virtualmario, C. Camp, J. D. Dworkin, and D. S. Bassett, "Gender diversity statement and code notebook v1.1," https://github.com/dalejn/cleanBib, October 2020.

In May 2024, we used Google Scholar's Cited by feature to retrieve every publication citing either anchor reference, with no restrictions on year, journal, or document type. After merging the two lists and removing duplicates, we obtained 510 unique candidate papers.

Two trained researchers then manually screened each candidate paper for the presence of a CDS. A paper was retained only if it contained a distinct, labeled section or subsection whose heading matched “Gender Diversity Statement,” “Citation Diversity Statement,” or a functionally equivalent variant (e.g., “Diversity in Citations”) and whose content followed the intent and structure of the cleanBib template. We excluded papers that (i) cited the anchor references without including any diversity statement, (ii) discussed gender diversity informally without a labeled section, or (iii) mentioned citation diversity only in acknowledgments or other unlabeled passages. No automated text-mining or keyword filtering was applied at this stage.

This procedure yielded 239 research articles with a qualifying CDS, published between January 2020 and May 2024.

### 2.2. Matching Design

To construct a comparison group, we drew on the Dimensions database, extracting all other research articles published between 2020 and 2024 in the same journals as the 239 CDS papers, excluding preprint repositories (e.g., bioRxiv, arXiv and PsyArXiv) to restrict the comparison pool to peer-reviewed venues. This yielded 420,515 candidate comparison papers. For each candidate article, we recorded four covariates: (i) the country of the first author’s primary affiliation, aggregated into geographic regions (e.g., North America, Europe, East Asia, South Asia); (ii) first-author gender, inferred from the first name using gender-guesser; (iii) research field, based on the two-digit (Division-level) ANZSRC Field of Research code; and (iv) team size (total number of listed authors), which we coarsened into quartiles prior to matching.

We then applied Coarsened Exact Matching (CEM) to pair each CDS paper with up to two comparison papers sharing identical coarsened covariate profiles on all four variables. After restricting to papers with complete reference records retrievable from Dimensions, the final matched dataset comprises 216 CDS papers (17,514 cited references) and 432 matched comparison papers (23,205 cited references), for a total of 648 papers and 40,719 cited references.

For each cited author, we inferred gender using gender-guesser and cultural origin using Ethnea (see Section S1 for tool descriptions and robustness checks with alternative inference tools). Main-text Fig. 1a illustrates the matching design.

### 2.3. Regression Specification

We estimated an ordinary least squares (OLS) regression of the form

$$\mathrm{CategoryRatio}_i = \beta_0 + \beta_1 \mathrm{CDS}_i + \gamma^{\top} C_i + \epsilon_i$$

where $\mathrm{CategoryRatio}_i$ denotes the proportion of cited authors in paper i belonging to a given gender or cultural-origin category (e.g., women, men, gender-blind; or English, Chinese, Japanese, Arab). $\mathrm{CDS}_i$ is a binary indicator equal to 1 if paper i includes a citation diversity statement and 0 for its matched comparison. $C_i$ represents a vector of control variables comprising publication year, number of co-authors (team size), total citation count (impact) entered as continuous covariates, and fixed effects for the first-ranked ANZSRC Field of Research code and the region of the first author's primary affiliation. Because field and region are also among the covariates used in the matching step, their inclusion as regression controls follows the standard "doubly robust" practice for coarsened exact matching, in which any residual within-stratum imbalance is absorbed by the corresponding fixed effects. The coefficient of interest, $\beta_1$, captures the average difference in citation share between CDS papers and their matched comparisons, holding the controls constant. Tables S2 and S3 report the resulting estimates for gender and cultural-origin categories, respectively.

## 3. Study 2 Method

Study 1 established that citation diversity statements are associated with increased recognition of Western women but reduced recognition of Eastern scholars at the population level. Study 2 moves to a controlled experimental setting to test whether this asymmetry is causally driven by the gender-feedback intervention itself. Study 2 tested whether an established gender-equity intervention, delivering feedback about the gender composition of a person's prior selections [15], disproportionately benefits women whose names are clearly gendered in English. The study was pre-registered on AsPredicted (https://aspredicted.org/py94-n9g5.pdf).

### 3.1. Experiment Data collection

To obtain a global sample of academics for our experiments, we drew on the Base of Human History and Talent (BHHT) public database of notable individuals with Wikipedia pages as of 2020 (https://medialab.github.io/bhht-datascape/). To fit the experimental setting, we restricted the pool to academics from three Western countries (United Kingdom, Italy, Germany) and three Eastern countries (China, South Korea, Thailand), yielding 4,417 academics. After cleaning and extracting each academic's name, gender, citizenship, education, and field of expertise, we performed stratified random sampling (stratified by gender and country) to select 10 women and 10 men from each country. This yielded a final master pool of 120 academics, balanced by gender (60 women, 60 men) and national origin (60 Western, 60 Eastern), spanning a range of academic disciplines and career stages.

For each of the 120 academics, we compiled a profile with five elements shown to participants: (i) first name, transliterated into English when necessary (e.g., Yong-phil, Luis); (ii) primary academic field (e.g., political science, mathematics); (iii) three research-interest keywords (e.g., systems theory; arithmetic geometry); (iv) the institution granting the academic's most recent degree (e.g., University of Chicago); and (v) the academic's professional title (e.g., Professor, Research Scientist, Academic Association President). These attributes were displayed in a structured Scholar Profile Table (see Fig. S7 for a screenshot).

For use in the feedback manipulations and subsequent analyses, we additionally coded four attributes for each scholar: gender (female or male), academic field category (Sciences, Social Sciences, Humanities and Arts, or Applied Fields), whether the scholar holds or held an administrative title (e.g., department chair, dean, institute director), and whether the degree-granting institution was public or private.

### 3.2. Participants and Procedure

We recruited 750 participants from the United States on Amazon's Mechanical Turk (MTurk) to complete a short survey. The sample included 46.0% men, 52.4% women, and 1.6% others, with an average age of 45.01 years ($SD = 12.82$). Participants self-identified as 74.2% White, 9.5% Black, 7.2% Hispanic, 8.3% Asian and 0.8% others.

Participants were truthfully informed that they would help select academics for a real Facebook advertisement featuring "outstanding academics" (Fig. S6). Each participant was shown a random sample of 30 academics drawn from the 120-scholar master pool: 24 men and 6 women, a ratio chosen to reflect the underrepresentation of women in academia. The sample spanned all six countries described in S3.1, and each scholar was presented via the Scholar Profile Table (Fig. S7). To incentivize careful selection, participants were offered a $5 bonus if their final set of chosen academics matched the set ultimately selected for the advertisement by a third party.

Participants first selected six academics to include in the Facebook campaign and then received one of two types of feedback at random. Half of participants were assigned to the no-gender-feedback condition. Here the feedback they received summarized their initial set of selectees without referencing their gender (it summarized the percentage of academics they had selected who (1) held an administrative position, (2) worked in the sciences, and (3) earned their PhDs at public universities). The other half of participants were assigned to the gender-feedback condition. In this condition, participants also received three pieces of feedback following their initial scientist selections: two randomly selected feedback elements from the three shown to participants in the no-gender-feedback condition, as well as the percentage of women among their initial selection. After receiving this condition-dependent feedback about the composition of their initial set of six academics to be included in a Facebook advertisement, all participants were invited to select a seventh and final academic to include in the Facebook campaign promoting the work of outstanding scholars. The identity of this final selection was the primary outcome of interest in our experiment.

### 3.3. Results

Following our pre-registered analysis plan, we ran ordinary least squares (OLS) regressions with robust (HC3) standard errors to predict the identity of the seventh academic selected. The primary predictor in every model was a binary indicator for assignment to the gender-feedback condition. Our primary analyses focused on two binary dependent variables: whether the seventh pick was (i) a woman from a Western country or (ii) a woman from an Eastern country. The main-text results for Study 2 are reported in Table 1 (main text) and visualized in Fig. 2b.

To formally test whether the feedback intervention prioritized Western over Eastern women, we re-estimated the two primary OLS regressions as seemingly unrelated regressions (SUR) and compared the two gender-feedback coefficients using a Wald test. The test confirmed that the treatment effect on Western-woman selection ($\beta = 0.1037$) was significantly larger than the effect on Eastern-woman selection ($\beta = 0.0268$; $\chi^2(1) = 4.450, p = 0.0349$).

## 4. Study 3 Method

Study 2 demonstrated that gender feedback disproportionately benefits Western women, whose names signal gender legibly in English. Study 3 extends this finding by asking whether shared cultural background between evaluators and scholars can close this gap, that is, whether Eastern evaluators, who are more familiar with Eastern names, respond differently to the feedback than Western evaluators do. Study 3 tested whether the asymmetry observed in Study 2 persists when evaluators come from different cultural backgrounds. By recruiting participants from both Western and Eastern countries, we examined whether shared cultural context between evaluators and scholars mitigates the legibility gap. The study was pre-registered on OSF (https://osf.io/j2ksg).

### 4.1. Participants and Procedure

We recruited 1,500 participants for this study, aiming for a balanced sample representing four national contexts as preregistered: 375 participants each from China, South Korea, Italy, and Germany. Participants were primarily recruited through the Prolific and Connect online platforms. Due to challenges in recruiting South Korean participants, 100 participants were recruited via emails sent by Yonsei University researchers to students at Yonsei University. Eligibility required fluency in the country's official language and either current residence or birth in that country. Participants were initially paid \$1, but to ensure balanced recruitment across regions, compensation was increased to \$2 after two weeks (63 participants from China, 71 from South Korea, 4 from Italy, and 24 from Germany were paid \$2). The final 100 South Korean participants recruited via Yonsei University received \$7. The final sample included 49.8% men, 48.9% women and 1.3% others, with an average age of 32.30 years ($SD = 10.32$). Participants self-identified as 47.2% White, 39.7% Asian, 11.7% Black, 1.2% Hispanic and 0.2% others.

Study 3 replicated the core design of Study 2 with minor adjustments to allow for cross-cultural comparison. Participants were shown a list of 24 academics, rather than 30, and asked to select six of them for another (real) Facebook campaign promoting outstanding scholars. Each list of available academics for the campaign included the same number of scholars from each of four countries: 6 from China, 6 from South Korea, 6 from Germany, and 6 from Italy, and these scholars were drawn from the same pool of 120 academics used in Study 2. Within each country, one woman and five men were randomly selected for inclusion in the candidate list, resulting in a total of 4 women and 20 men to be evaluated for inclusion in the Facebook advertisement.

After selecting six academics for inclusion in the Facebook advertisement, participants received feedback summarizing three descriptive statistics about their chosen academics. As before, those in the gender-feedback condition saw the percentage of women among their initial selections alongside two other pieces of feedback, while those in the no-gender-feedback condition viewed three other attributes (e.g. research field, all titles held, most recent degree institution) without gender information. All participants were then asked to choose a seventh and final academic to add to the Facebook advertisement from the original list of 24 candidates. Again, the identity of their seventh choice was the primary outcome of interest in our experiment.

### 4.2. Results

Following our pre-registered analysis plan, we first replicated the primary Study 2 analysis. Gender feedback significantly increased the selection of Western women ($\beta = 0.1360$, $p < 0.001$) but had no significant effect on the selection of Eastern women ($\beta = 0.0184$, $p =$ 0.233). A SUR model with a Wald test comparing these two coefficients confirmed that the gender feedback treatment had a significantly larger positive impact on the selection of Western women than Eastern women ($\chi^2(1) = 21.157$, $p < 0.001$).

We next tested whether this asymmetry was moderated by a participant's cultural background. Extending the previous SUR model, we included an interaction term between assignment to the gender-feedback condition and whether a participant was from a Western background, estimated separately for Western and Eastern female selection. Results from the interaction model (Table 2) showed the interaction between gender feedback and participant cultural background was significant for Western female selection ($\beta_{Western_woman:feedback \times western_participant} = 0.0750$, $p = 0.041$; Table 2 Model 2), indicating that Western participants responded somewhat more strongly to the intervention than Eastern participants when selecting Western women, but this difference was modest. Critically, no such interaction emerged for Eastern female selection ($\beta_{Eastern_woman:feedback \times western_participant} = 0.0188$, $p = 0.543$; Table 2 Model 4), indicating that Eastern participants were no more responsive than Western participants to the under-selection of Eastern women under the gender-feedback condition. A pre-registered Wald test comparing these two interaction terms finds they do not differ significantly ($\chi^2(1) = 1.204$, $p = 0.272$).

We conducted a follow-up analysis that decomposed our main effect by participant country of origin. Specifically, we tested whether participants recruited from each of the four countries—China, South Korea, Italy, and Germany—were more likely to select a female scholar from their own country when they received gender feedback. If shared cultural background were driving recognition, we would expect participants to disproportionately promote female scholars from their own country once aggregated gender feedback was provided.

For each country, we estimated a separate OLS regression predicting whether a participant's seventh and final selection was a female scholar from that country. Each model included an indicator for gender feedback, an indicator for whether the participant came from that same country ("Own-Country Participant"), and the interaction between the two, along with demographic controls (participant gender, race, and age) and incentive-level fixed effects. The interaction term is the focal coefficient: a positive and significant value would indicate that participants promoted female scholars from their own country specifically when given gender feedback about the gender composition of their initial selections.

The results, reported in Table S5, mirror our main findings. Gender feedback significantly increased the likelihood of selecting a female scholar from the two Western countries—Italy ($\beta = 0.070,\ p < 0.001$) and Germany ($\beta = 0.056, p < 0.001$)—but had no significant effect on the likelihood of selecting a female scholar from either Eastern country—China ($\beta = -0.005,\ p = 0.742$) or South Korea ($\beta = 0.017,\ p = 0.175$). Critically, the interaction between gender feedback and own-country participant status was non-significant in every model (China: $p = 0.268$; South Korea: $p = 0.918$; Italy: $p = 0.538$; Germany: $p = 0.510$). In other words, participants did not disproportionately select female scholars from their own country when given gender feedback about the gender composition of their selections, regardless of whether that country was Western or Eastern.

This pattern reinforces our central argument. The benefits of gender feedback accrued to female scholars from Western countries whether or not participants shared the scholar's national background, and did not accrue to female scholars from Eastern countries even when participants and scholars came from the same country. What determines whether feedback translates into recognition is therefore not the cultural proximity between evaluator and scholar, but whether the scholar's gender remains legible in the names that carry that signal.

Together, these findings tell a consistent story: the legibility gap is not a product of cultural unfamiliarity that dissolves when evaluators share a background with the scientists they are evaluating. Eastern participants conferred nearly as much benefit to Western women through the feedback as Western participants did, and neither group meaningfully increased their selection of Eastern women. Cultural background modestly attenuates the size of the advantage conferred to Western women, but does not close the gap — suggesting the mechanism is rooted in the linguistic structure of name-based gender inference rather than in culturally specific biases that familiarity could correct.

## 5. Figure Appendix

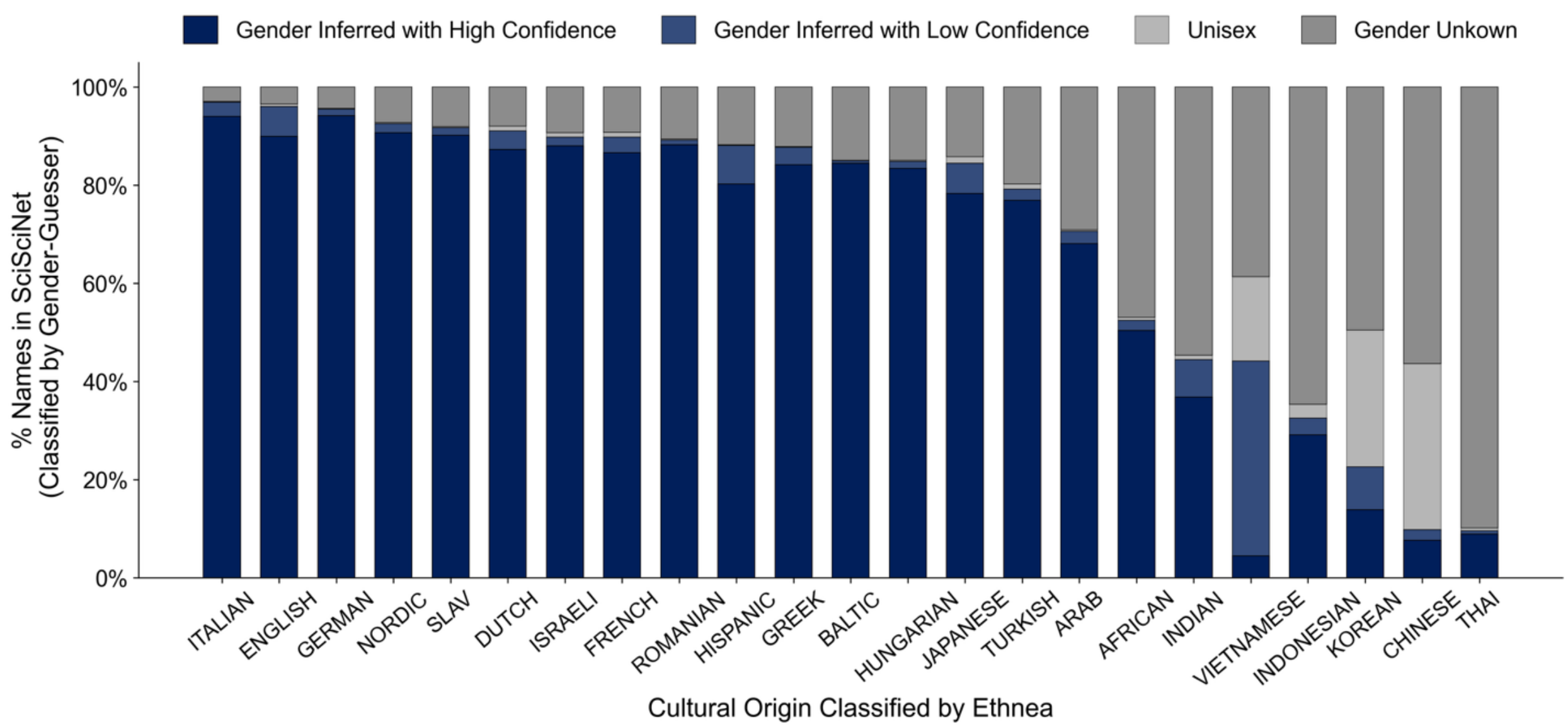

**Figure S1. Cross-cultural disparities in gender recognition by gender-guesser.** Proportion of names in the SciSciNet database (4.5 million scientists) classified by gender-guesser across 23 cultural groups identified by Ethnea. Bars show the share of names assigned to each output category: gender inferred with high confidence (female or male; dark blue), gender inferred with low confidence (mostly_female or mostly_male; light blue), unisex (andy; light gray), and gender-unknown (dark gray). Cultures are ranked left to right by the share of gender-signaling names. Western naming systems (e.g., Italian, English, German, Nordic) exhibit high recognition rates (~ 90%), while many Eastern naming systems (e.g., Chinese, Korean, Thai) exhibit low gender recognition rates (~ 10 %)

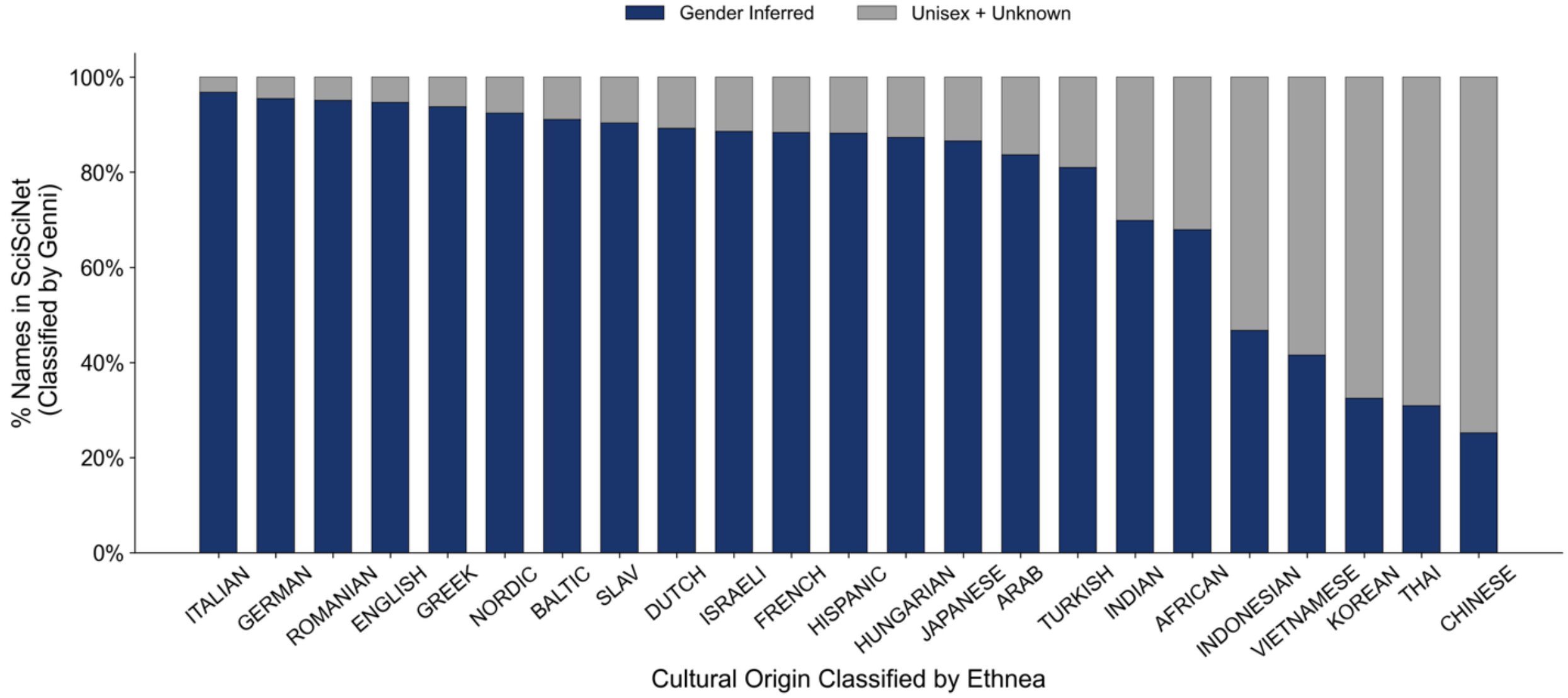

**Figure S2. Cross-cultural disparities in gender recognition by Genni.** Proportion of names in the SciSciNet database (4.5 million scientists) classified by Genni 2.0 across the same 23 cultural groups as in Fig. S1. Bars show the share of names assigned to each Genni output category: gender inferred (female or male; dark blue), or gender-blind (unisex or unknown; dark gray). Cultures are ranked left to right by the share of gender-signaling names (F + M). The cross-cultural pattern closely mirrors Fig. S1, with Western names recognized in the vast majority of cases and East and Southeast Asian names frequently returned as unknown.

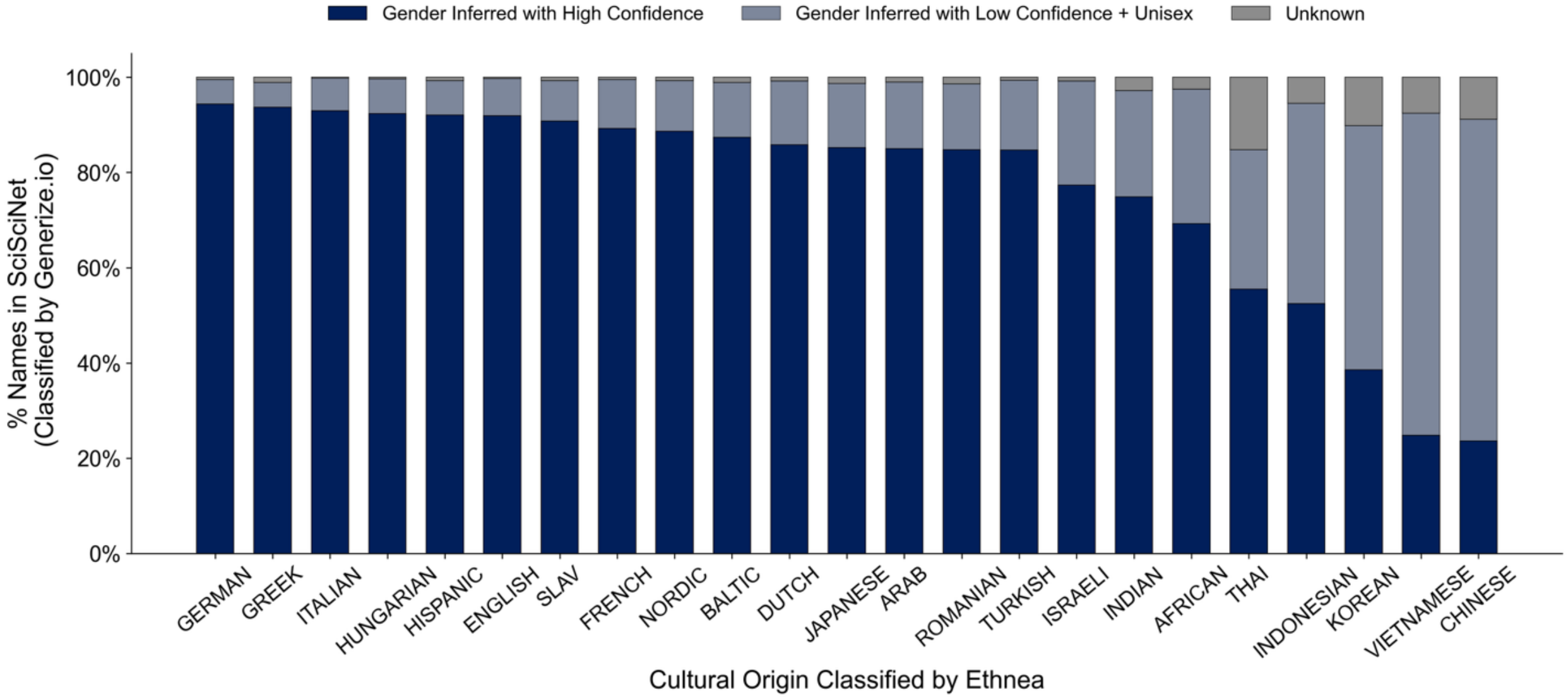

**Figure S3. Cross-cultural disparities in gender recognition by Genderize.io.** Proportion of names in the SciSciNet database (4.5 million scientists) classified by Genderize.io across the same 23 cultural groups as in Fig. S1. Bars show the share of names assigned to each output category based on the API's returned probability: gender inferred with high confidence ($\text{probability} > 0.95$; dark blue), gender inferred with low confidence or unisex ($0.5 \leq \text{probability} \leq 0.95$; light blue), and gender-unknown (name not found in the Genderize.io database; dark gray). Cultures are ranked left to right by the share of gender-signaling names with high confidence. The pattern closely parallels Figs. S1-S2: Genderize.io returns confident predictions for most Western names but low-confidence or unknown outputs for the majority of East Asian names.

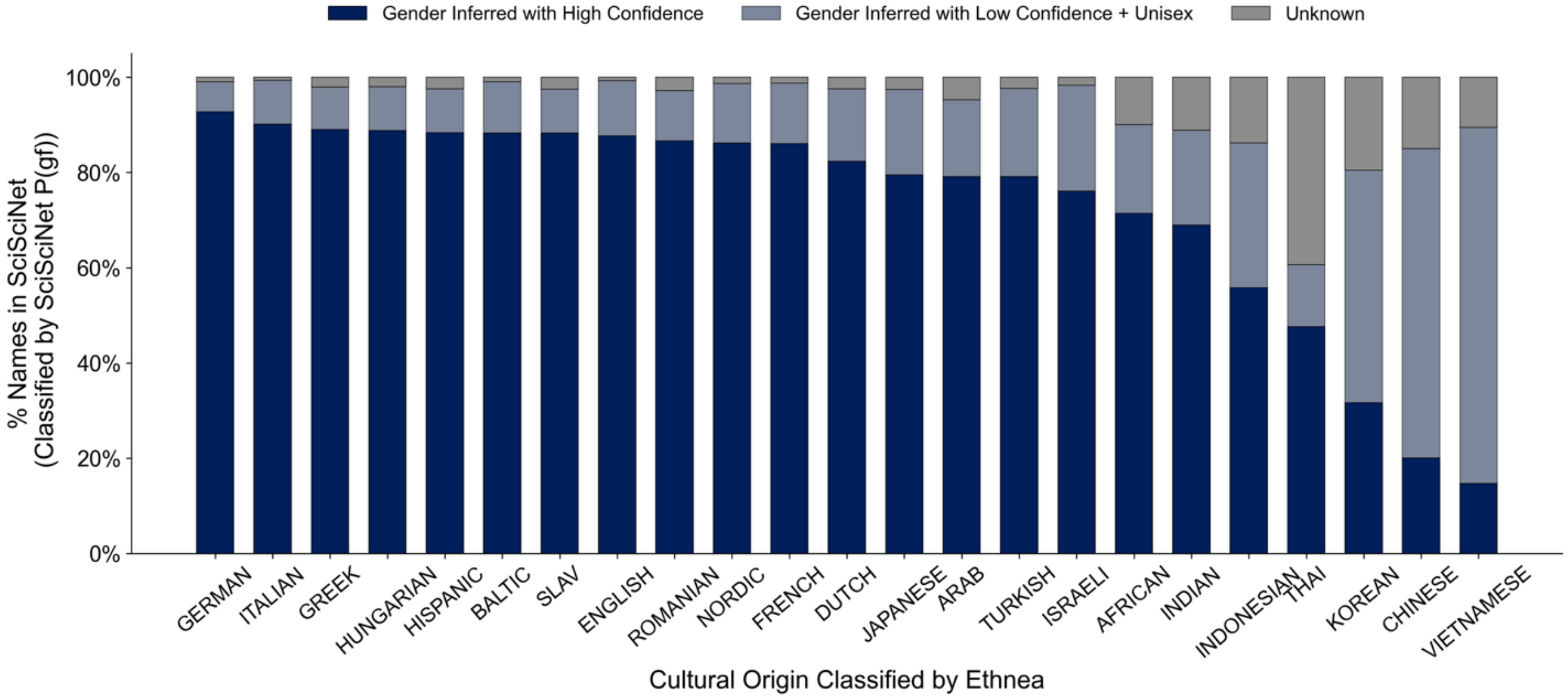

**Figure S4. Cross-cultural disparities in gender recognition using SciSciNet P(gf).** Proportion of names in the SciSciNet database for which cultural-consensus gender probability P(gf) is available, across the same 23 cultural groups as in Fig. S1. Bars show the share of names falling into each P(gf) range: gender inferred with high confidence ($P(gf) < 0.05$ or $> 0.95$; dark blue), gender inferred with low confidence or unisex ($0.05 \leq P(gf) \leq 0.95$; light blue), and gender unknown (dark gray). Cultures are ranked left to right by the share of gender-signaling names with high confidence. The pattern is consistent with Figs. S1-S3: P(gf) resolves gender confidently for the majority of Western names but frequently returns intermediate or unknown values for East Asian names.

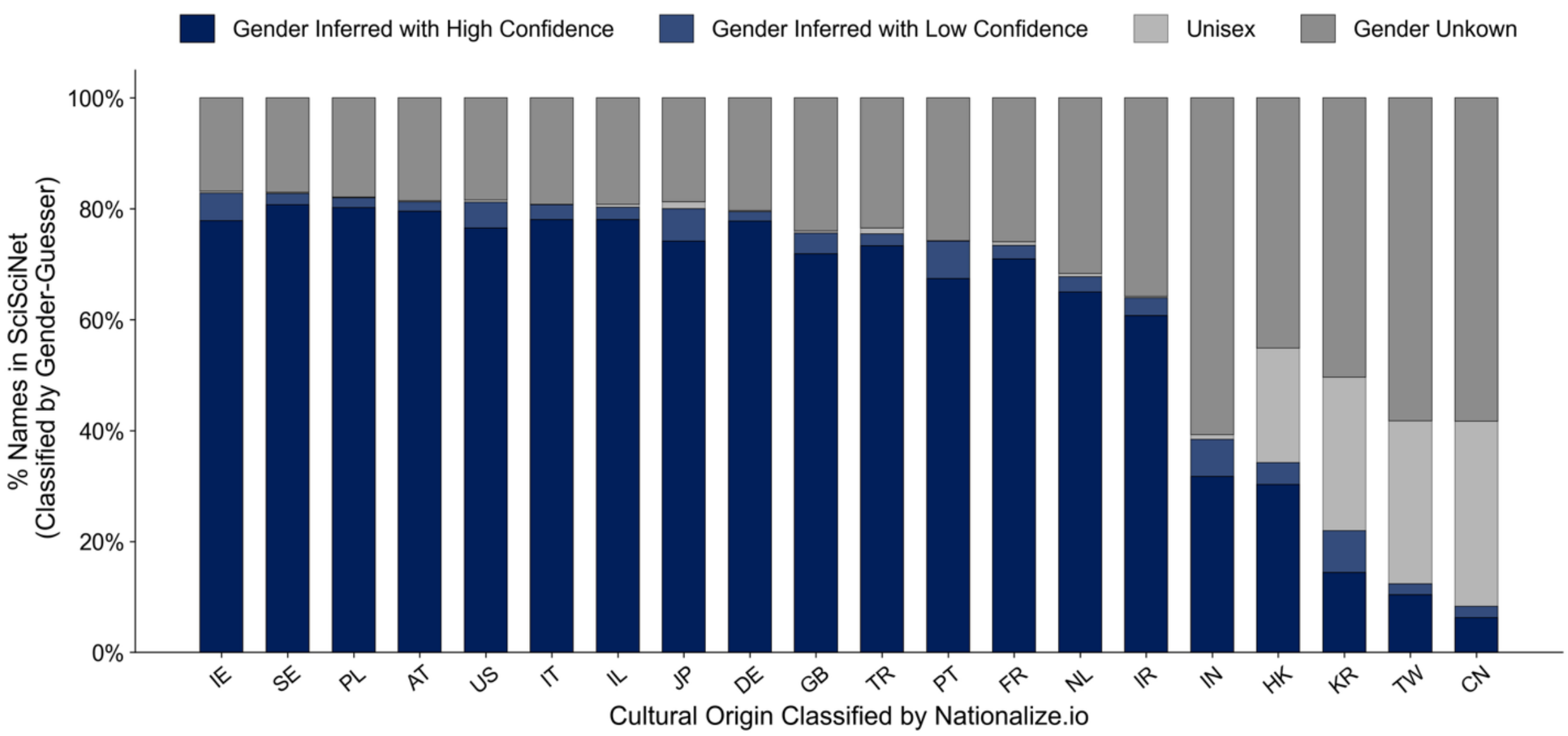

**Figure S5. Robustness of the gender legibility gap to an alternative cultural-origin classifier (Nationalize.io).** To verify that the cross-cultural pattern documented in Fig. S1 does not depend on our choice of cultural-origin classifier, we replicate the analysis using Nationalize.io instead of Ethnea to assign each name to a cultural group. This figure plots the top 20 cultural groups based on Nationalize.io's country predictions; gender is inferred from each author's first name using gender-guesser, as in Fig. S1. Cultural groups are ranked left to right by the share of gender-signaling names. The cross-cultural legibility gap persists under this alternative grouping: Western cultural groups show high recognition rates, whereas East and Southeast Asian groups show low recognition rates. This indicates that the pattern documented in Figs. S1-S4 reflects a genuine asymmetry in name-based gender inference, not an artifact of how cultural origin is assigned.

We are preparing a campaign on Facebook to highlight and promote a set of outstanding scientists. The purpose of the campaign is to increase public awareness about scientific endeavors.

The purpose of this HIT is to help us identify a group of scientists whose profiles you believe would resonate with the broad audience we are targeting with this campaign. We appreciate your assistance with this selection task in advance.

On the next page, you will see a list of 30 scholars from various scientific disciplines. We will ask you to select a subset of these scientists whom you believe would be of broad interest to the public and are therefore ideal to feature in our Facebook campaign.

To provide extra motivation to take this task seriously, we will send you a $5 bonus if your list is selected (there will be an eventual vote on all of the lists generated).

Here is a preview of the Facebook advertisement!

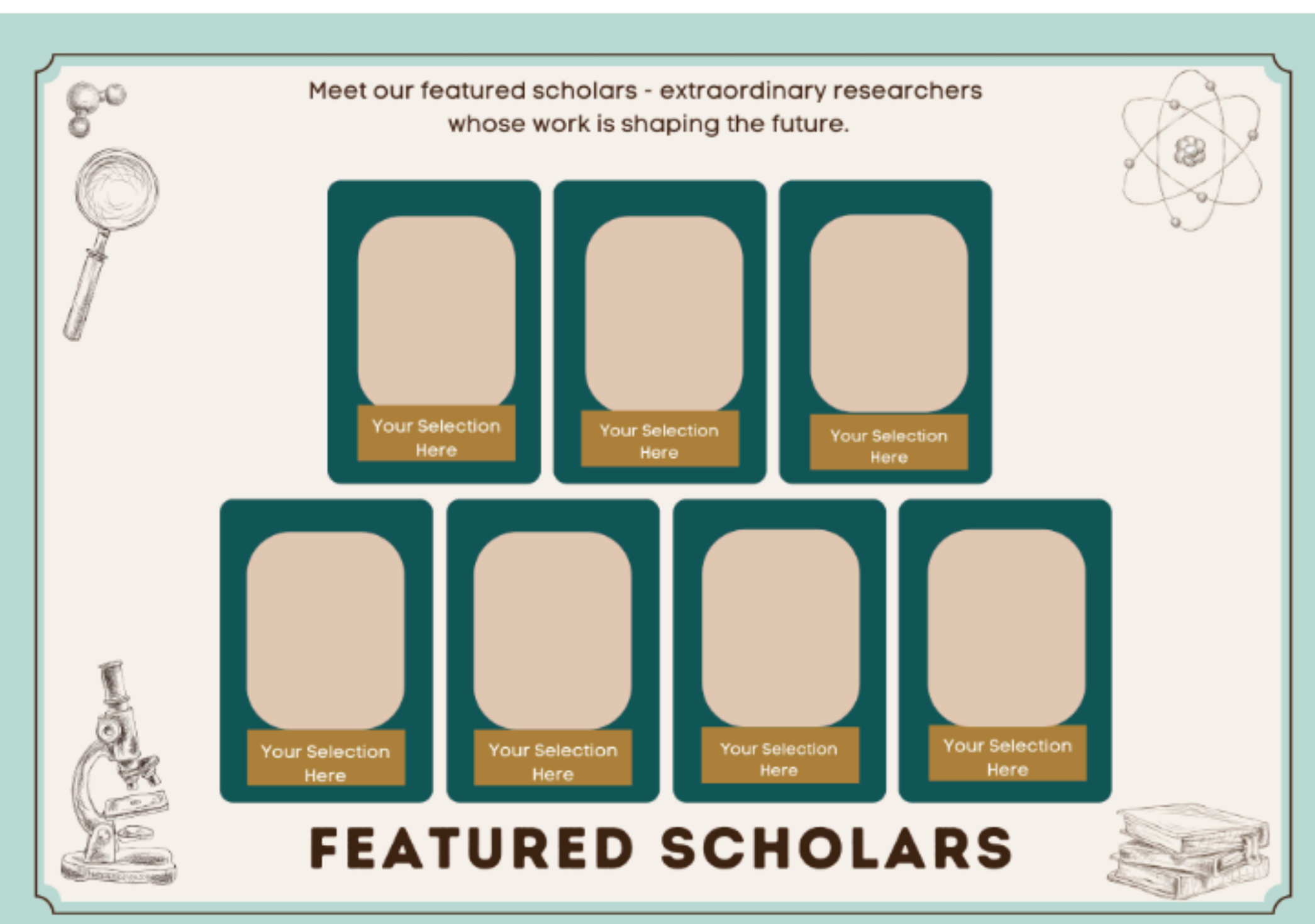

**Figure S6. Study introduction page.** Screenshot of the first page shown to participants in Studies 2 and 3, describing the Facebook campaign featuring outstanding research scholars, outlining the selection task, noting the bonus incentive, and previewing the layout of the final advertisement.

Please select the 6 professors you would like to include in the advertisement.

| **Name:** Ping | **Name:** Somsak | **Name:** Claus-Peter | **Name:** Shenggen | **Name:** Junliang | **Name:** Ghillean |
|---|---|---|---|---|---|
| **All Titles Held:** Scholar, President **Field:** Applied Fields **Research Interests:** civil law, legal reforms, legal education **Most Recent Degree from:** Moscow State University | **All Titles Held:** Scholar **Field:** Humanities and Arts **Research Interests:** political history, monarchy, democracy **Most Recent Degree from:** Monash University | **All Titles Held:** Scholar, Rector, Vice Rector **Field:** Humanities and Arts **Research Interests:** New Testament exegesis, Lukan theology, Passion narrative **Most Recent Degree from:** Philosophisch-Theologisches Studium Erfurt | **All Titles Held:** Scholar, Director **Field:** Sciences **Research Interests:** agriculture investments, rural poverty, infrastructure **Most Recent Degree from:** University of Minnesota | **All Titles Held:** Scholar **Field:** Humanities and Arts **Research Interests:** biblical studies, wisdom literature, theological interpretation **Most Recent Degree from:** Harvard University | **All Titles Held:** Scholar, Director, President **Field:** Sciences **Research Interests:** plant taxonomy, pollination ecology, Victoria amazonica **Most Recent Degree from:** Oxford University |
| **Name:** Willis | **Name:** Jong-sei | **Name:** Dong | **Name:** Mechtild | **Name:** Xinyuan | **Name:** Kwang-je |
| **All Titles Held:** Scholar **Field:** Humanities and Arts **Research Interests:** language teaching, discourse analysis, second language acquisition **Most Recent Degree from:** University of Bochum | **All Titles Held:** Scholar, FDA head **Field:** Sciences **Research Interests:** drug testing, biotechnology, protein microarray **Most Recent Degree from:** Johns Hopkins University | **All Titles Held:** Scholar, Director **Field:** Humanities and Arts **Research Interests:** unequal treaties, international relations, power dynamics **Most Recent Degree from:** East-West Institute of International Studies at Gordon College | **All Titles Held:** Scholar **Field:** Humanities and Arts **Research Interests:** life planning, gender relations, work-life balance **Most Recent Degree from:** Justus Liebig University Giessen | **All Titles Held:** Scholar, Director **Field:** Sciences **Research Interests:** cancer research, gene therapy, RNA structure-function **Most Recent Degree from:** Shanghai Institute of Biochemistry | **All Titles Held:** Scholar **Field:** Sciences **Research Interests:** particle physics, accelerator technology, synchrotron radiation **Most Recent Degree from:** University of Maryland |

**Figure S7. Scholar profile table.** Screenshot of the example of 12 scholar profile grid shown to participants during the initial six-scholar selection task in Studies 2 and 3, with each profile displaying the scholar's first name, academic field, research-interest keywords, the institution granting their most recent degree, and their professional title(s). Each participant was presented with a grid of 30 cards (24 men and 6 women, drawn randomly from the 120-scholar master pool described in Section S3.1) and instructed to review the profiles and select six scholars to nominate for the Facebook campaign. Study 3 used an analogous interface with 24 scholars rather than 30.

## 6. Table Appendix

| Domain | Platform / institution | Role of name-based gender inference | Source |
|---|---|---|---|
| Hiring & screening | SeekOut | Diversity filters let employers search specifically for women and other underrepresented candidates | help.seekout.com/help/Diversity-Filters |
| Hiring & screening | Gem | Diversity analytics track gender across the sourcing and hiring funnel | gem.com/solutions/diversity |
| Expert directories | Gage (500 Women Scientists) | Global directory of women and gender-diverse experts in STEMM for journalists and event organizers | gage.500womenscientists.org |
| Expert directories | SheSource (Women's Media Center) | Online database of women experts for journalists and media bookers | womensmediacenter.com/shesource |
| Organizational auditing | LinkedIn Talent Insights | Reports aggregate workforce gender composition, inferred where members do not self-identify | business.linkedin.com/talent-solutions/product-update/diversity-insights |
| Organizational auditing | Datapeople | Gender reports assign candidate gender from first names to audit the hiring funnel | help.datapeople.io/article/151-explore-gender-report |
| Official statistics | UK Intellectual Property Office | Infers inventor sex from names using birth-registry name–gender data | gov.uk (IPO, 2016) |
| Official statistics | USPTO | Attributes inventor gender via IBM Global Name Recognition and the WIPO gender-name dictionary | uspto.gov/.../Progress-and-Potential-2019.pdf |
| Official statistics | Science-Metrix, for U.S. NSF (NCSES) (2018) | Bibliometric gender indicators assigned with NamSor, building on the EU She Figures methodology across 41 countries | namsor.app/.../science-metrix_bibliometric_indicators...pdf |

Note: URLs are given in shortened form. All sources accessed July 2026.

**Table S1. Representative platforms, institutions, and tools that rely on name-based gender inference.** Across hiring, expert directories, organizational auditing, and official statistics, gender is routinely inferred from names in the absence of self-reported data. The list is illustrative rather than exhaustive.

| **Dependent variable:** | **Proportion of Women's Names in Citation List** | **Proportion of Men's Names in Citation List** | **Proportion of Gender-Blind Names in Citation List** |
|---|---|---|---|
| **Primary Predictor:** | 0.0709 *** | 0.0108 | -0.0860 *** |
| **Including CDS** | (0.0105) | (0.0103) | (0.0098) |
| Controls | Yes | Yes | Yes |
| Observations | 648 | 648 | 648 |
| R-Squared | 0.2890 | 0.0801 | 0.3297 |

**Table S2. OLS regressions estimating the association between citation diversity statements and citation composition by gender (Study 1).** OLS regressions estimating the relationship between inclusion of a citation diversity statement and the proportion of cited authors by gender (women's names, men's names, and gender-blind names). The key independent variable is a binary indicator for whether the paper includes a citation diversity statement. All models include controls for a paper's publication year, field of research, region of the first-ranked affiliated institution, number of authors, and total citation count. $+p < 0.10$; $* p < 0.05$; $** p < 0.01$; $*** p < 0.001$

| Dependent variable: Ethnicity composition | Primary Predictor: Including CDS | Control | Observations | R-Squared |
|---|---|---|---|---|
| ENGLISH | 0.0606 (0.0114) *** | Yes | 648 | 0.3154 |
| GERMAN | 0.0102 (0.0050) * | Yes | 648 | 0.1376 |
| ISRAELI | 0.0035 (0.0022) | Yes | 648 | 0.0834 |
| SLAV | 0.0020 (0.0038) | Yes | 648 | 0.1800 |
| FRENCH | 0.0019 (0.0034) | Yes | 648 | 0.0529 |
| DUTCH | 0.0014 (0.0025) | Yes | 648 | 0.1295 |
| TURKISH | 0.0011 (0.0012) | Yes | 648 | 0.0286 |
| ROMANIAN | -0.0001 (0.0005) | Yes | 648 | 0.0428 |
| INDONESIAN | -0.0002 (0.0001) + | Yes | 648 | 0.1332 |
| BALTIC | -0.0002 (0.0006) | Yes | 648 | 0.0517 |
| AFRICAN | -0.0005 (0.0012) | Yes | 648 | 0.0491 |
| VIETNAMESE | -0.0007 (0.0004) + | Yes | 648 | 0.0426 |
| HUNGARIAN | -0.0008 (0.0009) | Yes | 648 | 0.0497 |
| THAI | -0.0010 (0.0006) + | Yes | 648 | 0.0456 |
| NORDIC | -0.0013 (0.0029) | Yes | 648 | 0.0565 |
| GREEK | -0.0026 (0.0012) * | Yes | 648 | 0.0769 |
| KOREAN | -0.0039 (0.0021) + | Yes | 648 | 0.0532 |
| ITALIAN | -0.0058 (0.0033) + | Yes | 648 | 0.1044 |
| INDIAN | -0.0078 (0.0030) * | Yes | 648 | 0.1431 |
| ARAB | -0.0080 (0.0024) *** | Yes | 648 | 0.1554 |
| HISPANIC | -0.0099 (0.0043) * | Yes | 648 | 0.0702 |
| JAPANESE | -0.0101 (0.0028) *** | Yes | 648 | 0.1391 |
| CHINESE | -0.0316 (0.0071) *** | Yes | 648 | 0.1812 |

**Table S3. OLS regressions estimating the association between citation diversity statements and citation composition by ethnicity (Study 1).** OLS regressions estimating the relationship between inclusion of a citation diversity statement and the proportion of cited authors by ethnicity. The key independent variable is a binary indicator for whether the paper includes a citation diversity statement. All models include controls for a paper's publication year, field of research, region of the first-ranked affiliated institution, number of authors, and total citation count. $+p < 0.10$; $* p < 0.05$; $** p < 0.01$; $*** p < 0.001$.

| Tool name | Type | Version/Access | Input Format | Output Format |
|---|---|---|---|---|
| gender-guesser | Gender | gender-guesser 0.4.0 | First Name | "male", "female", "mostly_male", "mostly_female", "andy", or "unknown" |
| Genni | Gender | Genni 2.0 | Full Name | "M" (male), "F" (female), or "-" (unknown) |
| Genderize.io | Gender | Commercial API (Accessed Sep, 2024) | First Name | "male" or "female"; with probability (0-1) and data count |
| SciSciNet P(gf) | Gender | Public Dataset | AuthorId | Probability value P(gf) $\in [0, 1]$ |
| Ethnea | Cultural origin | Ethnea 2.0 | Full Name | Predefined ethnic groups; or "UNKNOWN" / "TOOSHORT" |
| Nationalize.io | Cultural origin | Commercial API (Accessed Sep, 2024) | Last Name | List of countries with associated probabilities summing to 1 |

**Table S4: Overview of gender and cultural-origin inference tools.** Summary of the six name-based inference tools used in this study, listing each tool's type (gender or cultural origin), version or access date, input format (first name, full name, last name, or AuthorId), and output format.

| Dependent variable: Identity of final academic selected: | China | South Korea | Italy | Germany |
|---|---|---|---|---|
| | Model 1 | Model 2 | Model 3 | Model 4 |
| **Intervention** (Feedback was provided on the % of academics who were women) | -0.0046 | 0.0168 | 0.0695 *** | 0.0559*** |
| | (0.0141) | (0.0124) | (0.0150) | (0.0158) |
| **Intervention x** Own-Country Participant | 0.0265 | -0.0025 | 0.0206 | 0.0215 |
| | (0.0239) | (0.0244) | (0.0333) | (0.0326) |
| Own-Country Participant | -0.0389 * | -0.0020 | -0.0013 | -0.0149 |
| | (0.0181) | (0.0176) | (0.0196) | (0.0190) |
| Controls | Yes | Yes | Yes | Yes |
| Observations | 1443 | 1443 | 1443 | 1443 |
| R-Squared | 0.0040 | 0.0025 | 0.0275 | 0.0160 |

**Table S5. Country-specific OLS regressions estimating the effect of gender feedback intervention on the selection of female scholars from each country (Study 3).** The table reports four OLS regressions, each predicting whether a participant's seventh and final scholar selection was a woman from a specific country: China (Model 1), South Korea (Model 2), Italy (Model 3), or Germany (Model 4). The key independent variable in every model is a binary indicator for random assignment to receive feedback on the percentage of women among the participant's initial six selections. "Own-Country Participant" is a binary indicator for participants residing in the same country as the scholars in the dependent variable (e.g., Chinese participants in Model 1), and its interaction with the feedback indicator tests whether participants disproportionately selected female scholars from their own country when given gender feedback intervention. All models include demographic controls (participant gender, race, and age) and incentive-level fixed effects; the analytic sample comprises the 1,443 participants who completed the task. Robust (HC3) standard errors appear in parentheses. $+p < 0.1$; $* \ p < 0.05$; $** \ p < 0.01$; $*** \ p < 0.001$.